\documentstyle[12pt,epsfig]{article}

\begin{document}
\baselineskip=23pt

\vspace{1.2cm}

\begin{center}
{\Large \bf  Nonmass Eigenstates of Fermion and Boson Fields}

\bigskip

Xin-Bing Huang\footnote{huangxb@shnu.edu.cn}\\
{\em Shanghai United Center for Astrophysics (SUCA),} \\
{\em   Shanghai Normal University, No.100 Guilin Road, Shanghai
200234, China}
\end{center}

\bigskip
\bigskip
\bigskip

\centerline{\large Abstract}

It appears natural to consider the four dimensional relativistic
massive field as a five dimensional massless field. If the fifth
coordinate is interpreted as the proper time, then the fifth moment
can be understood as the rest mass.  After introducing the rest mass
operator, we define the mass eigenstate and the nonmass eigenstate.
The general equations of spin-0, spin-$\frac{1}{2}$ and spin-1
fields are obtained respectively. It is shown that the Klein-Gordon
equation, the Dirac equation and the Proca equation describe the
mass eigenstates only. The rest mass of spin-$\frac{1}{2}$ field and
the rest mass squared of Boson fields are calculated. The $U(1)$
gauge field that couples to the nonmass eigenstates is studied
carefully, whose gauge boson can be massive.

\vspace{1.2cm}

PACS numbers: 12.15.Ff, 11.10.Kk, 12.60.Cn

\vspace{1.2cm}

\newpage

What is the origin of the rest mass? How to distinguish the massive
field and the massless field explicitly from the mathematical point
of view? What is the essential difference between the flavor
eigenstates and the mass eigenstates? All those problems are
fundamental, in which the problem of the rest mass has been studied
widely from different viewpoints. In the standard model of
electroweak interactions~\cite{gsw60,gm00}, the rest masses of
leptons and quarks originate from their Yukawa couplings with the
Higgs field~\cite{hig64}, and the mismatch between the flavor
eigenstates and the mass eigenstates of leptons and quarks is caused
by the Higgs interactions~\cite{xin04}. In the 5-dimensional
Kaluza-Klein theory~\cite{lm06,sp87, md91} and the 11-dimensional
string theory (or called M-theory)~\cite{pol98}, the quantum field
can obtain the rest mass via the compactification of extra
dimension. It is studied in Ref.\cite{fgl07} to consider the
4-dimensional relativistic particle as a 5-dimensional massless
particle and interpret the fifth coordinate as the particle's proper
time while the fifth moment can be understood as the mass. In this
letter, we use the proper time to define the rest mass operator and
discuss the nonmass eigenstates of Boson and Fermion fields.

Consider a Minkowskian spacetime with the following metric tensor
(covariant components)
\begin{equation}
\label{1001} \eta_{\mu\nu}= \left(
\begin{array}{cccc}
1 & 0 & 0 & 0
\\
0 & -1 & 0 & 0
\\
0 & 0 & -1 & 0
\\
0 & 0 & 0 & -1
\end{array}\right)
~,~~~~(\mu,\nu=0,1,2,3)~.
\end{equation}
In this letter we will use the contravariant three-vector
\begin{equation}
\label{1002}
x^{i}=\{x^{1},x^{2},x^{3}\}\equiv\{x,y,z\}~,~~~~(i=1,2,3)~,
\end{equation}
and four-vector
\begin{equation}
\label{1003} x^{\mu}=\{x^{0},x^{1},x^{2},x^{3}\}\equiv\{ct,x,y,z\}~,
\end{equation}
for the description of the spacetime coordinates, where the timelike
component is denoted as zero component. The proper time $s$ can be
given by
\begin{equation}
\label{1004+} s^{2}=x^{\mu}x_{\mu}=c^{2}t^{2}-x^{2}-y^{2}-z^{2}~,
\end{equation}
where $c$ is the speed of light in vacuum, which is invariant under
the Lorentz transformations. From (\ref{1004+}) one can acquire that
\begin{equation}
\label{1006} sds=ct d(ct)-xdx-ydy-zdz~,
\end{equation}
therefore
\begin{equation}
\label{1007} \frac{1}{c}\frac{\partial s}{\partial
t}=\frac{ct}{s}~,~~ \frac{\partial s}{\partial x}=-\frac{x}{s}~,~~
\frac{\partial s}{\partial y}=-\frac{y}{s}~,~~ \frac{\partial
s}{\partial z}=-\frac{z}{s}~,
\end{equation}
furthermore
\begin{equation}
\label{1008} \frac{\partial}{\partial
s}=\frac{s}{c^2t}\frac{\partial }{\partial t}-\frac{s}{x}
\frac{\partial }{\partial x}-\frac{s}{y}\frac{\partial}{\partial
y}-\frac{s}{z}\frac{\partial }{\partial z}={\bf
n}^{\mu}\partial_{\mu}~,~~\partial_{\mu}\equiv \frac{\partial
}{\partial x^{\mu}}~,
\end{equation}
where the contravariant vector ${\bf n}^{\mu}$ has been defined by
\begin{equation}
\label{1009} {\bf
n}^{\mu}=\left\{\frac{s}{ct},-\frac{s}{x},-\frac{s}{y},-\frac{s}{z}\right\}~.
\end{equation}

In order to discuss the rest mass we start by considering a free
particle with the relativistic relation
\begin{equation}
\label{1010}
\frac{E^2}{c^2}=p_{1}^{2}+p_{2}^{2}+p_{3}^{2}+m_{0}^{2}c^{2}~,
\end{equation}
here $m_{0}$ is the rest mass of the particle. Obviously the rest
mass plays the equal role with the component of momentum in above
equation, they therefore should have the similar quantization. In
elementary quantum mechanics~\cite{gri05}, the energy $E$ and the
component of momentum $p_{i}$ are quantized by
\begin{equation}
\label{1020} \hat{E}=i\hbar\frac{\partial}{\partial
t}~,~~~~\hat{p}_{i}=-i\hbar\frac{\partial}{\partial x^{i}}~,
\end{equation}
which are Hermitian operators, namely, $\hat{E}^{\dag}=\hat{E}$ and
$\hat{p}^{\dag}_{i}=\hat{p}_{i}$. If the rest mass is quantized by a
Hermitian operator also, then the real number $m_{0}$ in
(\ref{1010}) must be one of the eigenvalues of this Hermitian
operator. According to the relativistic relation (\ref{1010}) and
the operators of energy and momentum (\ref{1020}), we define the
rest mass operator as follows
\begin{equation}
\label{1030} \hat{m}=-i\frac{\hbar}{c}\frac{\partial}{\partial s}~,
\end{equation}
which is invariant under the Lorentz transfromations. It is easy to
prove that the rest mass operator is Hermitian if and only if $s$ is
timelike.

Assume that a set of eigenfunctions
$\sigma_j(x^{\mu})~(j=1,2,\cdot\cdot\cdot,n)$ constitutes an
$n$-dimensional complete Hilbert space, which are the eigenfunctions
of $\hat{m}$, and $m_j$ are the corresponding eigenvalues, which are
real and nonnegative--as the rest mass, of course, must be. We
reexpress this assumption in the mathematical language
\begin{equation}
\label{1060}
\hat{m}\sigma_j(x^{\mu})=-i\frac{\hbar}{c}\frac{\partial\sigma_j(x^{\mu})}{\partial
s}= m_{j}\sigma_j(x^{\mu})~,~~~~(j=1,2,\cdot\cdot\cdot,n)~.
\end{equation}
Above assumption shows that the mass eigenstate has been defined by
the eigenfunction of the rest mass operator. According to quantum
mechanics~\cite{gri05}, we can then define the nonmass eigenstate
$\sigma(x^{\mu})$ by
\begin{equation}
\label{1070} \sigma(x^{\mu})=\sum_{j=1}^{n}
a_j\sigma_j(x^{\mu})~,~~~~\sum_{j=1}^{n}a_j a_{j}^{*}=1~,
\end{equation}
where $a_j$ is complex and $a_{j}^{*}$ the complex conjugate of
$a_j$.

In quantum field theories, the mass operator had been introduced and
the nonmass eigenstates had also been used to denote the off-shell
states~\cite{wei95,str08}. But we give a quite different definition
for the rest mass operator. The nonmass eigenstate discussed by us
will devote it to a better understanding of particle mixing.

From elementary quantum mechanics it is known that the
schr{\"o}dinger equation corresponds to the non-relativistic energy
relation in operator form. we therefore replace the energy, the
momentum and the rest mass in the relativistic relation (\ref{1010})
by their corresponding operators (\ref{1020}) and (\ref{1030}) to
get a relativistic equation
\begin{equation}
\label{3010} \left(\frac{\partial^2}{c^2\partial
t^2}-\frac{\partial^2}{\partial x^2} -\frac{\partial^2}{\partial
y^2}-\frac{\partial^2}{\partial z^2}
-\frac{\partial^2}{\partial{s}^2}\right)\phi(x^{\mu})=0~,
\end{equation}
here $\phi(x^{\mu})$ is a nonmass eigenstate of the free scalar
fields. This equation is invariant under the Lorentz
transformations. For a mass-squared eigenstate $\phi(x^{\mu})$, we
obtain
\begin{equation}
\label{3031} \hat{m}^{2}\phi(x^{\mu})=
-\frac{\hbar^{2}}{c^{2}}\frac{\partial^2\phi(x^{\mu})}{\partial{s}^2}=m_{\phi}^{2}\phi(x^{\mu})~.
\end{equation}
Then (\ref{3010}) reduces to
\begin{equation}
\label{3040} \left(\frac{\partial^2}{c^2
\partial t^2}-\frac{\partial^2}{\partial x^2}
-\frac{\partial^2}{\partial y^2}-\frac{\partial^2}{\partial z^2}
+\frac{m_{\phi}^2c^2}{\hbar^2}\right)\phi(x^{\mu})=0~.
\end{equation}
The definition of the rest mass operator $\hat{m}$ shows that
(\ref{3031}) is the eigen equation of the operator $\hat{m}^2$.
Therefore, the function $\phi(x^{\mu})$ should be called the
mass-squared eigenstate of a scalar field. (\ref{3040}) is the
Klein-Gordon equation.

Assume that a set of mass-squared eigenstates $\phi_j(x^{\mu})$
given by (\ref{3031}) constitutes an $n$-dimensional complete
Hilbert space, where $m_j$ is the corresponding rest mass. Then the
nonmass eigenstate of spin-$0$ fields is generally given by
\begin{equation} \label{3170}
\phi(x^{\mu})=\sum_{j=1}^{n}a_{j}\phi_j(x^{\mu})
~,~~~~\sum_{j=1}^{n}a_{j}a_{j}^{*}=1~.
\end{equation}
Since $\phi_j(x^{\mu})$ are mass-squared eigenstates, one can only
obtain the square rest mass of a nonmass eigenstate defined by
(\ref{3170}) as follows
\begin{equation} \label{3180}
m^2=\sum_{j=1}^{n}a_{j}a_{j}^{*}m_{j}^{2}~.
\end{equation}

Following the historical approach of Dirac who, in 1928, obtained a
relativistic covariant wave equation for spin-$\frac{1}{2}$ field,
we give the relativistic covariant equation for the nonmass
eigenstate of spin-$\frac{1}{2}$ fields with a general potential
$V(x^{\mu})$
\begin{equation} \label{2070}
i\hbar\frac{\partial\psi(x^{\mu})}{\partial t}=\left[-i\hbar c
\hat{\alpha}^{i}\frac{\partial}{\partial x^{i}}-i \hbar c
\hat{\beta} \frac{\partial}{\partial s}+V\right]\psi(x^{\mu})~,
\end{equation}
where $\psi(x^{\mu})$ is a $4\times1$ matrix, and $\hat{\alpha}^{i},
\hat{\beta}$ are $4\times4$ Hermitian matrices defined by Dirac. One
can prove the covariance of this equation by noticing that the
operator $\frac{\partial}{\partial s}$ is invariant under the
Lorentz transformations.

For a mass eigenstate $\psi(x^{\mu})$, we get
\begin{equation}
\label{2071}
\hat{m}\psi(x^{\mu})=-i\frac{\hbar}{c}\frac{\partial\psi(x^{\mu})}{\partial
s}=m_{\psi}\psi(x^{\mu})~,
\end{equation}
and
\begin{equation}
\label{2110} \left(i\hbar\gamma^{\mu}\frac{\partial}{\partial
x^{\mu}}-m_{\psi}c\right)\psi(x^{\mu})=\frac{V}{c}\gamma^{0}\psi(x^{\mu})~,
\end{equation}
here we have adopted the definitions of
$\gamma^{\mu}=\{\gamma^{0},\gamma^{1},\gamma^{2},\gamma^{3}\}$ and
$\gamma^{0}=\hat{\beta},~\gamma^{i}=\hat{\beta}\hat{\alpha}^{i}$.
From above analysis, we can draw a conclusion that the mass
eigenstate of the spin-$\frac{1}{2}$ field is described by the
well-known Dirac equation (\ref{2110}).

From (\ref{2070}) we find that the Lagrange density of a nonmass
eigenstate of free spin-$\frac{1}{2}$ particles has the
form\footnote{In this letter ${\cal L}_{1n}$ denotes the Lagrange
density of one nonmass eigenstate and ${\cal L}_{1m}$ denotes the
Lagrange density of one mass eigenstate.}
\begin{eqnarray}\label{2150}
 {\cal L}_{1n}={\bar
{\psi}}(x^{\mu})\left(i\hbar c \gamma^{\mu}\frac{\partial}{\partial
x^{\mu}}+i\hbar c \frac{\partial}{\partial s}\right)
{\psi}(x^{\mu})~.
\end{eqnarray}
$\bar{\psi}\equiv{\psi}^{\dag}\gamma^{0} $ is called the spinor
adjoint to $\psi$. The variation of above ${\cal L}_{1n}$ with
respect to ${\bar {\psi}}(x,z)$ yields the general equation for a
nonmass eigenstate of free spin-$\frac{1}{2}$ Fermions
\begin{eqnarray}\label{2151}
\left(\gamma^{\mu}\frac{\partial}{\partial
x^{\mu}}+\frac{\partial}{\partial s}\right) \psi(x^{\mu})=0~.
\end{eqnarray}
The above equation is (\ref{2070}) with $V=0$. The Dirac equation in
the models of flat $1+4$ dimensional spacetime is generally given
by~\cite{sp87,md91,ich02,wu08}
\begin{eqnarray}\label{2153}
\left(i\hbar\gamma^{\mu}\frac{\partial}{\partial
x^{\mu}}+i\hbar\gamma^{5}\frac{\partial}{\partial x^{5}}-mc\right)
\Psi(x,x^{5})=0~,
\end{eqnarray}
where the metric of 5-dimensional spacetime is of the signature
$(+,-,-,-,-)$ and $\gamma^{5}=-\gamma_{5}\equiv
-i\gamma^{0}\gamma^{1}\gamma^{2}\gamma^{3}$. Obviously (\ref{2151})
can be treated as the massless 5-dimensional Dirac equation.

We can obtain the Lagrange density of a mass eigenstate of the free
spin-$\frac{1}{2}$ particle from (\ref{2150}), that is
\begin{eqnarray}\label{2160}
 {\cal L}_{1m}={\bar
{\psi}}(x^{\mu})\left(ic\hbar\gamma^{\mu}\frac{\partial}{\partial
x^{\mu}}-m_{\psi}c^{2}\right)
\psi(x^{\mu})~,~~~~\bar{\psi}\equiv{\psi}^{\dag}\gamma^{0}~.
\end{eqnarray}

Assume that a set of mass eigenstates $\psi_j(x^{\mu})$ given by
(\ref{2071}) constitutes an $n$-dimensional complete Hilbert space,
where $m_j$ is the corresponding rest mass. Then the nonmass
eigenstate of spin-$\frac{1}{2}$ fields must be
\begin{equation} \label{2170}
\psi(x^{\mu})=\sum_{j=1}^{n}a_{j}\psi_{j}(x^{\mu})
~,~~~~\sum_{j=1}^{n}a_{j}a_{j}^{*}=1~.
\end{equation}
According to quantum mechanics, the rest mass of nonmass eigenstate
$\psi(x^{\mu})$ given by (\ref{2170}) is therefore of the form
\begin{equation} \label{2180}
m=\sum_{j=1}^{n}a_{j}a_{j}^{*}m_{j}~.
\end{equation}
In a word, we have given the equation for a nonmass eigenstate of
spin-$\frac{1}{2}$ fields and the rest mass of a nonmass eigenstate.

The gauge field theories~\cite{gm00,sch96,fra08,cl84,ryd96} tell us
a good method to introduce the spin-1 fields in our framework. In
quantum field theories, we are familiar with the electromagnetic
field, which is a massless $U(1)$ gauge field. From the Lagrange
densities (\ref{2150}) and (\ref{2160}), we find that both of them
admit the introduction of a $U(1)$ gauge field. Since ${\cal
L}_{1m}$ can be treated as the Lagrangian of a free charged
spin-$\frac{1}{2}$ particle in quantum field theory, introducing a
$U(1)$ gauge field into the Lagrange density (\ref{2160}) will
directly give an electromagnetic field. The total Lagrangian
\begin{equation}
\label{4001}
 {\cal L}_{1mt}={\bar
{\psi}}\left(ic\hbar\gamma^{\mu}\frac{\partial}{\partial
x^{\mu}}-e\gamma^{\mu}A_{\mu}-m_{\psi}c^{2}\right)
\psi-\frac{1}{4}F_{\mu\nu}F^{\mu\nu}
\end{equation}
is invariant under the following local $U(1)$ transformation
\begin{equation}
\label{4010}
{\psi}^{\prime}(x^{\mu})=e^{i\theta(x^{\mu})}\psi(x^{\mu})~,~~~~A^{\prime}_{\mu}=A_{\mu}-
\frac{\hbar c}{e}\frac{\partial \theta(x^{\mu})}{\partial x^{\mu}}~,
\end{equation}
where $\theta(x^{\mu})$ is a function of $x^{\mu}$, and $A_{\mu}$ is
the electromagnetic field. The electromagnetic field strength tensor
$F_{\mu\nu}=\frac{\partial A_{\nu}}{\partial x^{\mu}}-\frac{\partial
A_{\mu}}{\partial x^{\nu}}$ is invariant under above transformation
as well.

From now on we will pay more attention to introducing a $U(1)$ gauge
field into the Lagrange density ${\cal L}_{1n}$. To make the gauge
invariance explicit, we formally define
\begin{equation}
\label{4302} x^{4}=-x_{4}=s~,~~~~\gamma^{4}=-\gamma_{4}={\bf 1}~.
\end{equation}
Thus the Lagrange density of a nonmass eigenstate of free
spin-$\frac{1}{2}$ fields (\ref{2150}) becomes
\begin{eqnarray}\label{4303}
 {\cal L}_{1n}={\bar
{\psi}}(x^{\mu})\left(ic\hbar\gamma^{\alpha}\frac{\partial}{\partial
x^{\alpha}}\right) \psi(x^{\mu})~,~~~~\alpha=0,1,2,3,4~.
\end{eqnarray}
Let us multiply the nonmass eigenstate $\psi(x^{\mu})$ by a local
phase $e^{i\Theta(x^{\mu})}$, that is
\begin{equation}
\label{4300}
\psi^{\prime}(x^{\mu})=~e^{i\Theta(x^{\mu})}\psi(x^{\mu})~.
\end{equation}
We can reexpress the above equation as doing a $U(1)$ transformation
on $\psi(x^{\mu})$ because the phase factor $e^{i\Theta(x^{\mu})}$
is an element of $U(1)$ group. So naturally
\begin{equation}
\label{4301}
{\bar{\psi}}^{\prime}(x^{\mu})=~e^{-i\Theta(x^{\mu})}{\bar{\psi}}(x^{\mu})~.
\end{equation}
The crucial result is that the total Lagrange density
\begin{eqnarray}\label{4401}
 {\cal L}_{1nt}={\bar
{\psi}}(x^{\mu})\left(ic\hbar\gamma^{\alpha}\frac{\partial}{\partial
x^{\alpha}}-g\gamma^{\alpha}{\bf A}_{\alpha}(x^{\mu})\right)
\psi(x^{\mu})-\frac{1}{4}{\bf F}_{\alpha\beta}(x^{\mu}){\bf
F}^{\alpha\beta}(x^{\mu})
\end{eqnarray}
is invariant under a group of local gauge transformations, given by
(\ref{4300}), (\ref{4301}) and
\begin{equation}
\label{4311} {\bf A}^{\prime}_{\alpha}(x^{\mu})={\bf
A}_{\alpha}(x^{\mu})- \frac{\hbar c}{g}\frac{\partial
\Theta(x^{\mu})}{\partial x^{\alpha}}~,
\end{equation}
$g$ in above equations is the coupling constant. The strength tensor
of $U(1)$ gauge field is of the form
\begin{equation}
\label{4321} {\bf F}_{\alpha\beta}(x^{\mu})=\frac{\partial {\bf
A}_{\beta}(x^{\mu})}{\partial x^{\alpha}}-\frac{\partial {\bf
A}_{\alpha}(x^{\mu})}{\partial x^{\beta}}~,
\end{equation}
which is invariant under the transformations of (\ref{4300}),
(\ref{4301}) and (\ref{4311}) as well. Now we will prove that ${\bf
A}_{\alpha}(x^{\mu})$ is a four dimensional covariant vector. To do
this, we decompose ${\bf A}_{\alpha}(x^{\mu})$ into two parts
\begin{equation}
\label{4351} {\bf A}_{\alpha}\equiv \{{\bf A}_{\mu},{\bf A}_{s}\}~.
\end{equation}
Considering (\ref{1008}) and (\ref{4401}) together, we get
\begin{equation}
\label{4352} {\bf A}_{s}=\frac{s}{ct}{\bf A}_{0}-\frac{s}{x} {\bf
A}_{1}-\frac{s}{y}{\bf A}_{2}-\frac{s}{z}{\bf A}_{3}={\bf
n}^{\mu}{\bf A}_{\mu}~.
\end{equation}
Combining (\ref{4311}) with (\ref{4352}) gives
\begin{equation}
\label{4354} {\bf A}_{s}^{\prime}={\bf n}^{\mu}{\bf
A}_{\mu}^{\prime}={\bf n}^{\mu}\left({\bf A}_{\mu} - \frac{\hbar
c}{g}\frac{\partial \Theta}{\partial x^{\mu}}\right)={\bf
A}_{s}-\frac{\hbar c}{g}\frac{\partial \Theta}{\partial s}~.
\end{equation}
Therefore ${\bf A}_{\alpha}$ is a four-vector and (\ref{4352}) is in
accordance with (\ref{4354}).

Compare the Lagrange density (\ref{4401}) with (\ref{4001}), we find
that the $U(1)$ gauge field ${\bf A}_{\alpha}(x^{\mu})$ can be
treated as a 5-dimensional Maxwell's electromagnetic field.
Therefore
\begin{equation}
\label{4600} \frac{\partial {\bf F}_{\alpha\beta}(x^{\mu})}{\partial
x^{\gamma}}+\frac{\partial {\bf F}_{\gamma\alpha}(x^{\mu})}{\partial
x^{\beta}}+\frac{\partial {\bf F}_{\beta\gamma}(x^{\mu})}{\partial
x^{\alpha}}=0~,
\end{equation}
and
\begin{equation}
\label{4605} \frac{\partial {\bf F}_{\alpha\beta}(x^{\mu})}{\partial
x_{\alpha}}={\bf J}_{\beta}(x^{\mu})~,
\end{equation}
where ${\bf J}_{\beta}(x^{\mu})$ is the 5-dimensional current.

Substituting (\ref{4321}) into (\ref{4605}) we find that ${\bf
A}_{\alpha}(x^{\mu})$ satisfies
\begin{equation}
\label{4606} \frac{\partial}{\partial x_{\alpha}}\frac{\partial {\bf
A}_{\beta}(x^{\mu})}{\partial x^{\alpha}}-\frac{\partial}{\partial
x^{\beta}}\frac{\partial {\bf A}_{\alpha}(x^{\mu})}{\partial
x_{\alpha}}={\bf J}_{\beta}(x^{\mu})~.
\end{equation}
We may now make use of the freedom (\ref{4311}) and choose a
particular $\Theta(x^{\mu})$ so that the transformed ${\bf
A}_{\alpha}(x^{\mu})$ satisfies the following gauge condition:
\begin{equation}
\label{4608} \frac{\partial {\bf A}_{\alpha}}{\partial
x_{\alpha}}=0~.
\end{equation}
In this special ``choice of gauge", (\ref{4606}) becomes
\begin{equation}
\label{4610} \frac{\partial}{\partial x_{\alpha}}\frac{\partial {\bf
A}_{\beta}(x^{\mu})}{\partial x^{\alpha}}={\bf J}_{\beta}(x^{\mu})~.
\end{equation}

In vacuo, there are no current, namely ${\bf J}_{\beta}(x^{\mu})=0$,
(\ref{4610}) reduces to
\begin{equation}
\label{4350} \left(\frac{\partial^2}{c^2\partial
t^2}-\frac{\partial^2}{\partial x^2} -\frac{\partial^2}{\partial
y^2}-\frac{\partial^2}{\partial z^2}
-\frac{\partial^2}{\partial{s}^2}\right){\bf A}_{\alpha}=0~.
\end{equation}
This is the general equation of a nonmass eigenstate of free spin-1
fields. Under this special choice of gauge, let us consider a
mass-squared eigenstate of ${\bf A}_{\mu}$, which is given by
\begin{equation}
\label{4355} \hat{m}^{2}{\bf
A}_{\mu}=-\frac{\hbar^{2}}{c^2}\frac{\partial^2 {\bf
A}_{\mu}}{\partial{s}^2}=m_{{\bf A}}^2{\bf A}_{\mu}~,~~~~ m_{{\bf
A}}\ge 0~.
\end{equation}
Then (\ref{4350}) reduces to
\begin{equation}
\label{4360} \left(\frac{\partial^2}{c^2
\partial t^2}-\frac{\partial^2}{\partial x^2}
-\frac{\partial^2}{\partial y^2}-\frac{\partial^2}{\partial z^2}
+\frac{m_{{\bf A}}^2c^2}{\hbar^2}\right){\bf A}_{\mu}=0~.
\end{equation}
Hence we have shown that the mass-squared eigenstates (or simply
called ``the mass eigenstates") of free vector field satisfies Proca
equation (\ref{4360}). Therefore the gauge boson of $U(1)$ gauge
field that couples to the nonmass eigenstates can be massive.

Assume that a set of mass-squared eigenstates $\left[{\bf
A}_{\mu}(x^{\mu})\right]_{j}$ given by (\ref{4355}) constitutes an
$n$-dimensional complete Hilbert space, where $m_j$ is the
corresponding rest mass. Then the nonmass eigenstate of vector
fields is expressed by
\begin{equation} \label{4700}
{\bf A}_{\mu}(x^{\mu})=\sum_{j=1}^{n}a_{j}\left[ {\bf
A}_{\mu}(x^{\mu})\right]_{j}~,~~~~\sum_{j=1}^{n}a_{j}a_{j}^{*}=1
\end{equation}
It is proved that $\left[{\bf A}_{\alpha}(x,z)\right]_{j}$ are
mass-squared eigenstates, one can calculate the square rest mass of
the nonmass eigenstate defined by (\ref{4700}), namely
\begin{equation} \label{4800}
m^2=\sum_{j=1}^{n}a_{j}a_{j}^{*}m_{j}^{2}~.
\end{equation}
Obviously the vector field has the same rest mass formula as that of
the scalar field.

It is well-known that the mass term in Lagrangian of a charged
particle must be invariant under the Lorentz transformations and the
local gauge transformations, therefore, for a spin-$\frac{1}{2}$
nonmass eigenstate who couples with a $U$(1) gauge field, the rest
mass operator must be invariant not only under the Lorentz
transformations but also under the $U$(1) gauge transformations.
Hence, the rest mass operator of $\psi(x^{\mu})$ in Lagrangian
(\ref{4401}) should be
\begin{equation}
\label{mass} \hat{M}=-i\frac{\hbar}{c}\left(\frac{\partial}{\partial
s}+i\frac{g}{\hbar c}{\bf A}_{s}\right)~.
\end{equation}
One can easily prove that $\bar{\psi}(x^{\mu})\hat{M}\psi(x^{\mu})$
is invariant under the Lorentz transformations and the local $U$(1)
gauge transformations.

In the Standard Model, the Yukawa interactions of the quarks with
the Higgs condensate cause the mismatch between the flavor
eigenstates $d_{k}^{\prime}$ and the mass eigenstates $d_{l}$,
namely $d_{k}^{\prime}\equiv\sum_{l}V_{kl}d_{l}, ~k,l=1,2,3,$ and
$V$ is the Cabibbo-Kobayashi-Maskawa mixing
matrix~\cite{cab63,km73}, which is unitary $V^{\dag}V=1$. In our
framework, the mass eigenstates of the quarks must be written as
$d_{l}$, where $m_{l}$ is the corresponding mass of the quark. Then
the nonmass eigenstates of the quarks are of the form
\begin{equation} \label{2190}
d_{k}^{\prime\prime}=\sum_{l=1}^{3}V_{kl}d_{l}~.
\end{equation}
In the forthcoming papers~\cite{hua09b,hua09c} we will prove that
the nonmass eigenstates play an important role in constructing an
electroweak model without Higgs mechanism.

By interpreting the proper time as the fifth coordinate, we define
the operator of the rest mass and give the concepts of mass
eigenstate and nonmass eigenstate. The general equations for nonmass
eigenstates of free spin-0, spin-$\frac{1}{2}$ and spin-1 fields are
obtained. It is found that there are two kinds of $U(1)$ gauge
fields: The $U(1)$ gauge field of first kind merely couples to mass
eigenstates, in which the gauge boson is massless. The second kind
of $U(1)$ gauge field couples to nonmass eigenstates, whose gauge
boson may be massive.

{\bf Acknowledgement:} I am grateful to Prof. Chao-Jun Feng and
Prof. Dao-Jun Liu for their enlightening discussions. It is very
important that Prof. Chao-Jun Feng indicated that the rest mass
operator could be well defined by proper time without introducing an
extra dimension.


\end{document}